\documentstyle[11pt,aasms,tighten,flushrt]{article}
\def \yrs{~\rm{yrs}}

\def \pc{~\rm{kpc}}
\begin{document}


\title{
THE `SECOND PARAMETER': A MEMORY FROM THE \\
GLOBULAR CLUSTER FORMATION EPOCH}

\author{ 
Noam Soker and Ron Hadar$^1$}
\affil{
Department of Physics, University of Haifa at Oranim\\
Oranim, Tivon 36006, ISRAEL \\
soker@physics.technion.ac.il \\      
$^1$ permanent position: West Valley High School, Israel}


\bigskip
\bigskip
\centerline {\bf ABSTRACT}
\begin{abstract}

 We study the correlations between the distribution of stars on the
horizontal branch (HB), the HB morphology, with some other properties
of globular clusters (GCs) in a large sample of GCs.
 We strengthen previous results that a general correlation
exists only between HB morphology and metallicity.
Correlations with other properties, e.g., central density and
core radius, exist only for GCs within a narrow metallicity range.
 We conjecture that the lack of correlations with {\it present}
properties of GCs (besides metallicity), is because the variation
of the HB morphologies between GCs having similar metallicities
is caused by a process, or processes, whose effect was determined
at the {\it formation time} of GCs.
 This process (or processes) is historically termed the
`second parameter', metallicity being the `first parameter'.
We then argue that the `planet second parameter' model, where the
presence of planets and to a lesser degree brown dwarfs and low
mass main sequence stars is the `second parameter', fits
this conjecture.
This is because the processes which determine the presence
of planets and their properties occur during the formation
epoch of the star and its circumstellar disk.

\end{abstract}



\section{INTRODUCTION}

 Evolved sun-like stars which burn helium in their cores occupy the
horizontal branch (HB) in the Hertzsprung-Russel (HR) diagram. 
 The distributions of stars on the HB (the HB morphology)
differ substantially from one globular cluster (GC) to another. 
 It has long been known that metallicity is the main, but not sole,
factor which determines the HB morphology
(for a historical review see, e.g., Rood, Whitney, \& D'Cruz 1997;
Fusi Pecci \& Bellazzini 1997).
 The other factor (or factors) which determines the HB morphology is
termed the `second parameter'.
 In recent years it has become clear that mass loss on the red giant branch
(RGB) is closely connected with the second parameter, in the sense that the
second parameter should determine the HB morphology by regulating
the mass loss on the RGB  (Dorman, Rood, \& O'Connell 1993; 
D'Cruz {\it et al.} 1996;  Ferraro {\it et al.} 1998;
Whitney {\it et al.} 1998; Catelan 2000).
D'Cruz {\it et al.} (2000) suggest that the the amount of mass
lost on the RGB can even determine the gaps observed in the
distribution of stars along the HB (e.g., in NGC 2808,
Bedin {\it et al.} 2000).

 But what is the process which determines the total mass that is
lost by a star on the RGB, and why does it operate differently
in different GCs?
 What other effects does this process have on the HB morphology?
 Is the variation of mass loss rate by itself sufficient to
explain the entire `second parameter'?
 Sweigart \& Catelan (1998; also Sweigart 1999; Moehler,
Sweigart, \& Catelan 1999; and Moehler {\it et al.} 2000), for example,
claim that mass loss on the RGB by itself cannot be the second
parameter, and it should be supplied by another process, e.g.,
rotation, or helium mixing, which requires rotation as well. 
 They term the addition of such a process a `noncanonical scenario'.
Behr {\it et al.} (2000) find the second parameter problem to be
connected with rotation, and note that single star evolution cannot
explain the observed rotation of HB stars, even when fast core
rotations are considered.
 Soker \& Harpaz (2000) analyze the angular momentum evolution from
the main sequence to the HB.
 Using rotation velocities for stars in the GC M13, they find that
the required angular momentum for the fast rotators is up to
$1-3$ orders of magnitude larger than that of the sun.
 Soker \& Harpaz (2000) argue that their results show that the
fast rotating HB stars have been probably spun-up by planets,
brown dwarfs, or low-mass main sequence stars, while they evolved
on the RGB.
 They further argue that these angular momentum considerations
support the `planet second parameter' model.
 In this model (Soker 1998) the `second parameter' process
is the interaction of RGB stars with low-mass companions, in most cases
gas giant planets, and in a minority of cases with brown dwarfs or
low-mass main sequence stars.
 The masses and initial orbital separations of the planets
(or brown dwarfs or low-mass main sequence stars) form a rich
spectrum of different physical parameters, which manifests itself in
the rich varieties of HB morphologies observed in the different
globular clusters, e.g., different values of HBR, gaps and tails.

 One of the common approaches in the search for the second parameter
is to look for correlations between the HB-morphology and other
GC properties other than metallicity.
 Although some correlations for a limited number of GCs were found,
they are weak and are limited to only a fraction of all GCs.
 In many cases the quantity that is used to characterize the
HB-morphology is HBR, which is defined as
$HBR \equiv (B-R)/(B+V+R)$.
Here $V$ is the number of HB-stars within the instability strip,
and $R$ and $B$ are the number of stars redder and bluer, respectively,
than the instability strip.
 This quantity strongly correlates with metallicity,
which is the `first parameter' (e.g., Lee 1990).
Fusi Pecci {\it et al.} (1993) and Buonanno {\it et al.} (1997)
find some correlations between the HB-morphology and $\rho_0$,
the central density of GCs (both papers use a different quantity
from HBR to characterize the HB morphology).
 Both papers find the correlation to be weak and/or limited to
a small fraction of all GCs in their sample.
 Fusi Pecci {\it et al.} (1993) find correlations with both the
ellipticity and total GC luminosity.
 Again, these correlations were weak, and significant only for GCs
within a narrow range of metallicity.
 The lack of clear correlations between the HB morphology and GCs'
properties (besides metallicity), which holds for most GCs,
constrain any second parameter process.
 The limited correlation with the central density of GCs, for example,
may hint at binary mergers and/or tidal stripping of RGB stars'
envelope as the second parameter mechanism.
 However, there are several problems with this idea
(e.g., Landsman {\it et al.} 1996; Rood {\it et al.} 1997; Rood 1998;
Ferraro {\it et al.} 1997, 1998), e.g., the presence of blue HB stars
in the core of the galaxy M32 (Brown {\it et al.} 2000),
where the stellar density is much lower than that in GCs.
 Soker (1998) argues that Rood's (1998) three points against stellar
binary scenarios do not hold for the planet second parameter
model.

 In the present paper we propose that the lack of clear correlations
with {\it present} GCs' properties (besides metallicity) results
from the fact that the second parameter reflects GCs' properties during
the {\it formation time} of GCs.
We then discuss how the planet second parameter model fits
this claim.
 In $\S 2$ we present some correlations between HB morphology (HBR)
and other GCs properties. These will serve our discussion
in $\S 3$.  Our summary is in $\S 4$.

\section{CORRELATIONS WITH GLOBULAR CLUSTERS' PROPERTIES}

 As the basis for the discussion to follow in the next section, we
present some correlations, or non-correlations, of the HB morphology
with some GC properties.
 Most of these were found in the past, but it will be useful to
present them here in somewhat different forms.
 For this we use the tables of Milky Way GCs compiled by
Harris (1996; available via
http://physun.physics.mcmaster.ca/Globular.html, updated June 1999).
 With the HB morphology parameter defined in the previous section:
$HBR \equiv (B-R)/(B+V+R)$.
 In Figure 1 we present the well known correlation (e.g., Lee 1990)
between HBR and metallicity (the `first parameter')
for 114 GCs, as indicated in the figure itself
(for 33 out of the 147 there were no HBR and/or
metallicity given by Harris 1996).
 From this correlation we can distinguish three main parts:
low metallicity GCs, [Fe/H]$\lesssim -1.8$, all have a large
fraction of blue HB stars, i.e., HBR$>0.1$; high metallicity GCs,
[Fe/H]$\gtrsim -0.9$, all have a low fraction of blue stars;
the middle metallicity group,
$-1.8 \lesssim$[Fe/H]$\lesssim -0.9$, have a spread in HBR
over the entire possible range.
 This last group is the focus of the rest of this section.
The correlation between the HB morphology and metallicity points
to the sensitivity of the star location on the HB to its metallicity,
higher metallicity stars being redder.
 This is because the metallicity has another effect, which makes
stars bluer on the HB, by which higher metallicity RGB stars reach
larger radii on the RGB tip, hence a higher mass loss rate is expected
(D'Cruz {\it et al.} 2000).
Higher mass lost on the RGB means a bluer HB star.
Despite this effect, low metallicity GCs have bluer HB morphologies,
which means that the mass loss effect on the RGB plays a secondary role.
 As mentioned earlier, it was suggested (e.g.,
Dorman, Rood, \& O'Connell 1993; Catelan 2000), that the total mass
lost on the RGB is connected with the second parameter.
The question is what determines the amount of mass lost on the RGB.
 The discussion in the first section of this paper leads us to support the
view that a single mechanism causes both higher mass loss rate on the RGB
and a `noncanonical effect', e.g., helium mixing plus rotation
(Sweigart \& Catelan 1998; Sweigart 1999;
Moehler {\it et al.} 1999).
 Following works by Soker (1998), Siess \& Livio (1999),
Behr {\it et al.} (2000) and Soker \& Harpaz (2000), we take 
this mechanism to be induced rotation in RGB stars,
mainly by planets.
 Rotation along the RGB may increase the total mass loss, increase
the core mass at helium flash (Mengel \& Gross 1976), and mix helium
to the envelope (Sweigart \& Catelan 1998, and references therein).

 Next we search for a correlation with the central density of the
cluster $\rho_0$, as given in units of solar luminosity per cubic
parsec.
 Such a correlation was noted before (e.g., Fusi Pecci {\it et al.}
1993; Buonanno {\it et al.} 1997; ).
Instead of doing it for the entire sample, we divide the entire sample
of 147 GCs into four equal groups according to metallicity, with dividing
values at [Fe/H]=$-1.675$, $-1.42$, and $-0.78$, as indicated by
vertical lines in Figure 1.
 In Figure 2 we present the results for the three lower metallicity
quarters (again, not all GCs have HBR values, hence the number of
GCs in each quarter is below 36 GCs).
 We find correlation between HBR and $\rho_0$ {\it only in
the second quarter}, where the metallicity is in the range of
$-1.675 <~$[Fe/H]$~< -1.42$.
Interestingly, the only GC which is not along the diagonal strip
of the correlation in the second quarter, NGC 7492, has a very
elliptical shape, with minor to major axes ratio of $b/a= 0.76$
(White \& Shawl 1987).
 We can still keep the correlation in the second quarter if we add
the next five GCs below metallicity of [Fe/H]$=-1.675$.
 One of these, NGC 6144, is to the upper left of the correlation,
but it has high ellipticity, $b/a=0.75$, similar to NGC 7492.
 The correlation is also kept if we add the next five GCs above
metallicity of [Fe/H]$=-1.42$.
 Therefore the correlation exists in the metallicity range of
$-1.75 \lesssim$[Fe/H]$\lesssim -1.35$.
 Adding more GCs on either sides makes the correlation worse
(see Fig. 4 below), hence there is no general correlation between
HB morphology and GCs' densities.
 It is also interesting to note that the third quarter,
with $-1.42 <~$[Fe/H]$~< -0.78$, shows no correlation
with $\rho_0$, despite the spread of HBR on the entire possible
range (-1 to +1).
 This demonstrates the points mentioned in the first section,
which were found by many authors previously:
some correlations hold for a small fraction of all GCs,
but besides metallicity no other property holds correlation for
all GCs.

 Noting the high ellipticity of NGC 7492, and the claim for
correlation with ellipticity in the metallicity range of
$-1.9 \leq $[Fe/H]$\leq -1.4$ (Norris 1983), we turn to examine
correlations with the ellipticity of GCs, taken from White \& Shawl
(1987; note that not all GCs have measured ellipticity).
 Again we divide the sample into four groups according to metallicity.
In Figure 3 we present the results for the second and third quarters.
 For the second quarter a weak correlation exist in the sense that
high ellipticity GCs (low b/a) have high values of HBR.
 However, the third quarter does not possess this correlation.
Moreover, if we remove the three GCs with [Fe/H]$<-1.4$,
marked with squares in the figure, we find a weak {\it opposite}
correlation, as did Fusi Pecci {\it et al.} (1993).
 We conclude that the positive correlation of blue HB morphology
with ellipticity  (Norris 1983) is limited to a very narrow range of
metallicity, and it is definitely not a general correlation.

 To further demonstrate the differences between the metallicity groups
in figure 4 we plot the value of HBR versus the core radius
$r_c$, for four different metallicity ranges.
 In the lower panel we present the correlation for the second quarter
we used in Figures 2 and 3; the metallicity range is
$-1.675<~$[Fe/H]$~< -1.42$, with 30 GCs.
 In the upper panel we extend the metallicity band to
$-1.75<~$[Fe/H]$~<-1.35$, where the correlation still holds;
there are 36 GCs marked with asterisks.
 As shown on the same panel, the correlation no longer exists
for a lower metallicity band
$-1.95<~$[Fe/H]$~< -1.75$ (15 GCs marked with circles) 
and for a higher metallicity band
$-1.35<~$[Fe/H]$~< -1.15$ (9 GCs marked with squares).  
 Again, we see that the correlation, in the sense that GCs with
small core radius tend to have high HBR, holds only for GCs in
the metallicity range $-1.75 \lesssim$[Fe/H]$\lesssim -1.35$.
 We note that similar correlations, although less prominent,
can be found in this metallicity band between the HB morphology and
other GC properties, e.g., absolute luminosity of the cluster
and the central concentration $c$ (Fusi Pecci {\it et al.} 1993),
half mass radius $r_h$, tidal radius $r_t$, $r_t/r_c$,
$r_h/r_c$, and the core relaxation time $t_c$.

\section{DISCUSSION: PROPERTIES AT THE FORMATION EPOCH}

 We conjecture that the lack of any clear correlation with {\it present}
properties of GCs, besides metallicity, hints that the second parameter
effect was determined at the formation time of the GC.
 Norris (1983) already suggested that the second parameter is the
rotation of single stars, and that the rotations of stars in a
GC are determined at the formation time of the GC.
 He based his proposal on the correlation he found between
HB morphology and the ellipticity of GCs (but see previous section),
which he assumed is linked to the rotations of single stars.
 The rotation rate of a young star is determined by the interaction
between the protostellar disk and the young star.
The properties of the protostellar disk are influenced by the
environment; the star is too small to be
influenced directly by the environment. 
 Therefore, the formation of planets (and brown dwarf and low mass
main sequence stars) at several AU from the central star is much
more sensitive to the environment than the rotation of the
young star.
 The lack of clear general correlation with GCs' ellipticity (see
previous section), and the higher sensitivity of the disk 
than that of the young star embedded inside it
to the environment make the presence of binary companions,
substellar or stellar, a better candidate for the second parameter
than the rotations of single stars.
 We argue that the discussion above further supports the
`planet second parameter' model (Soker 1998; see first section here).

 The environment can influence the planet formation efficiency
and the subsequent planet properties in many ways, most of
which are poorly understood.
 In a high density region close stars can cause disturbances in
the outskirts of the protoplanetary disk, causing more material
to stream to the inner region.
 This may enhance planet formation, via a strong perturbation
in the gravitational instability process (Boss 1998), and by
increasing the disk mass, making planet formation more efficient in
the planetesimal accretion process (e.g., Kenyon {\it et al.} 1999).
 But the stellar density around a star during the formation of
its planets in a GC has no simple relation to its present density,
core radius, and other parameters.
 The young GC goes through a violent relaxation phase, in which
dense regions are formed and destroyed (van Albada 1982), and
whose evolution depends on the initial conditions prior to
the violent relaxation (May \& van Albada 1984).
 Interestingly, the typical violent relaxation time in GCs,
found by scaling the results of van Albada (1982) to
initial mass of $10^6 M_\odot$ and radius of $10 \pc$,
is within an order of magnitude of the typical life time
of circumstellar disks, $\sim 10^7 \yrs$.
 This means that on average a star goes through dense regions 
no more than a few times during the formation time of planets,
on the assumption that planet formation and GC formation occur
at the same time; the exact number of times is strongly dependent
on the initial condition of the GC.
 Both processes of planet and GC formation during the
relevant epochs are too poorly known to say more here.
 Finally, we note that the formation of planets in the planetesimal
accretion process can be very efficient (e.g., Kenyon \& Luu 1999),
but it still depends on the metallicity of the circumstellar disk.
 This may be one of the reasons, but not the only one,
behind the different correlations in the different
metallicity classes (Figs. 2-4).

\section{SUMMARY}

The goal of the present paper was to present further arguments
in favor of the `planet second parameter' model.
In this model the main factor after metallicity
(the `first parameter') of the `second parameter' of the HB morphology
is the presence of planets, and to lesser degree brown dwarfs and
low mass main sequence companions. 
 The model was suggested by Soker (1998), with further supporting
arguments, mainly angular momentum considerations, in
Soker \& Harpaz (2000).
 In the present paper we discussed the correlations, or lack of
correlations, between GC properties and the HB morphology (Figs. 1-4).
These correlations have been long known, but we have presented them in
a way appropriate for our purpose.
Our conclusions are as follows. 
\newline
(1) In addition to being the main factor which determines HB
morphology (Fig. 1), the `first parameter', metallicity, also
plays a role in determining whether other correlations exist.
That is, some correlations exist in a specific metallicity range
but not in others.
\newline
(2) GCs' parameters for which partial correlations (i.e., only
in a narrow band of metallicity) hold are the cluster central
density and similar related parameters, such as the core radius.
The correlation with ellipticity is very weak in this metallicity
range, and changes its behavior for a slightly higher metallicity
(Fig. 3).
\newline
(3) Despite the partial correlations,  no general correlation
exists with any present property of GCs (besides metallicity).
We presented the lack of general correlation for central density,
ellipticity, and core radius, Figures 2-4, respectively,
but checked it for other properties.
\newline
(4) We conjectured that the HB morphology is correlated with a
property (or properties) at the {\it formation time} of GCs.
That is, we argued that the second parameter process (or processes)
is fixed at the formation time of the GCs, and not later in
the evolution.
 A similar idea, but with respect to single star rotation,
was put forward by Norris (1983).
\newline
(5) We argued that the `planet second parameter' goes along with
this conjecture, since the processes which determine the presence
of planets and their properties occur during the formation
epoch of the star and its circumstellar disk.
 The dependence of planet formation efficiency on metallicity
can be one of the factors behind the existence of
correlations only in a narrow band of metallicity.
 At this point we cannot expand on this point because of
lack of knowledge on the formation of both GCs and planets. 

In the present study we used only the HBR value to characterized
the HB morphology.
When more data become available correlations with other
properties of the HB should be sought.
 Ferrao {\it et al.} (1998), for example, note that most high
density GCs with metallicity of [Fe/H]$\sim -1.5$ have blue HB
tails.
Any model for the second parameter should eventually account for
these fine details, e.g., tails and gaps, in the HB morphology.
In the planet second parameter model, extreme blue HB stars
are attributed to interaction with massive planets, brown dwarfs,
or low mass main sequence stars (Soker \& Harpaz 2000)

\acknowledgments
This research was supported in part by grants from the 
Israel Science Foundation and the US-Israel Binational
Science Foundation.

{\bf FIGURE CAPTIONS}

\noindent {\bf Figure 1:}
 The HB morphology, as given by HBR, versus metallicity.
Values are taken from Harris (1996; updated June 1999 in
http://physun.physics.mcmaster.ca/Globular.html).
 HBR is defined as $HBR \equiv (B-R)/(B+V+R)$, where $V$ is the
number of HB-stars within the instability strip, and $R$ and $B$
are the number of stars redder and bluer, respectively,
than the instability strip.
 The three vertical lines are the boundaries between the four quarters
of 147 GCs, determined by equal number of GCs in each group.
 These groups are used in Figures 2-4.
 Only $114$ GCs are shown, since some of the 147 GCs given
by Harris (1996) have no HBR value.

\noindent {\bf Figure 2:}
 The HBR value (see Fig. 1) versus the central density
(in solar luminosity per cubic parsec).
Values are from Harris (1996, updated June 1996).
 The number of GCs and the metallicity range are given inside each panel.
The lower, middle, and upper panels are for the first (lower), second,
and third metallicity quarters marked on Figure 1, respectively. 

\noindent {\bf Figure 3:}
 The HBR value (see figure 1) versus ellipticity, as given
by the ratio of minor to major axis of the GC.
HBR values are from Harris (1996, updated June 1996),
and ellipticities are from White \& Shawl (1987).
 The number of GCs and the metallicity range are given inside each panel.
The lower and upper panels are for the second and third metallicity
quarters marked on Figure 1, respectively.
 The three GCs marked by squares in the upper panel have metallicity
close to that of the lower panel, $-1.42 <~$[Fe/H]$~< -1.4$.

\noindent {\bf Figure 4:}
 The HBR value (see Fig. 1) versus the core radius (in parsec).
 Values are from Harris (1996, updated June 1996).
The lower panel is for the second metallicity quarter (see Fig. 1).
In the upper panel the metallicity range of the second quarter has
been extended to $-1.75 <~$[Fe/H]$~< -1.35$, and the number of GCs
in that range is now 36 (marked with asterisks).
 The weak correlation seen in the lower panel still exists.
 For a lower metallicity band
$-1.95 <~$[Fe/H]$~< -1.75$, 15 GCs marked by circles,
and for a higher metallicity band
$-1.35 <~$[Fe/H]$~< -1.15$, 9 GCs marked by squares,
the correlation no longer exists.
            
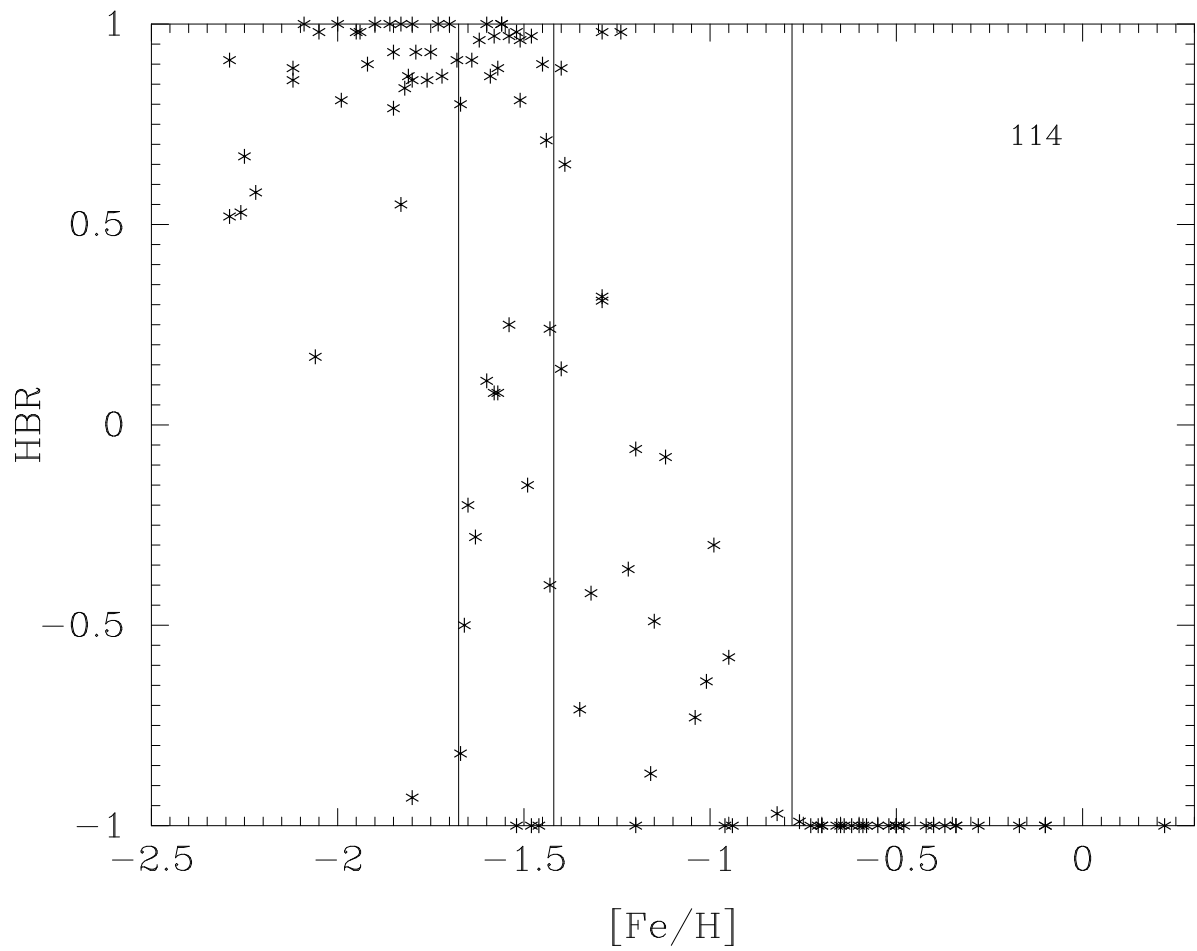
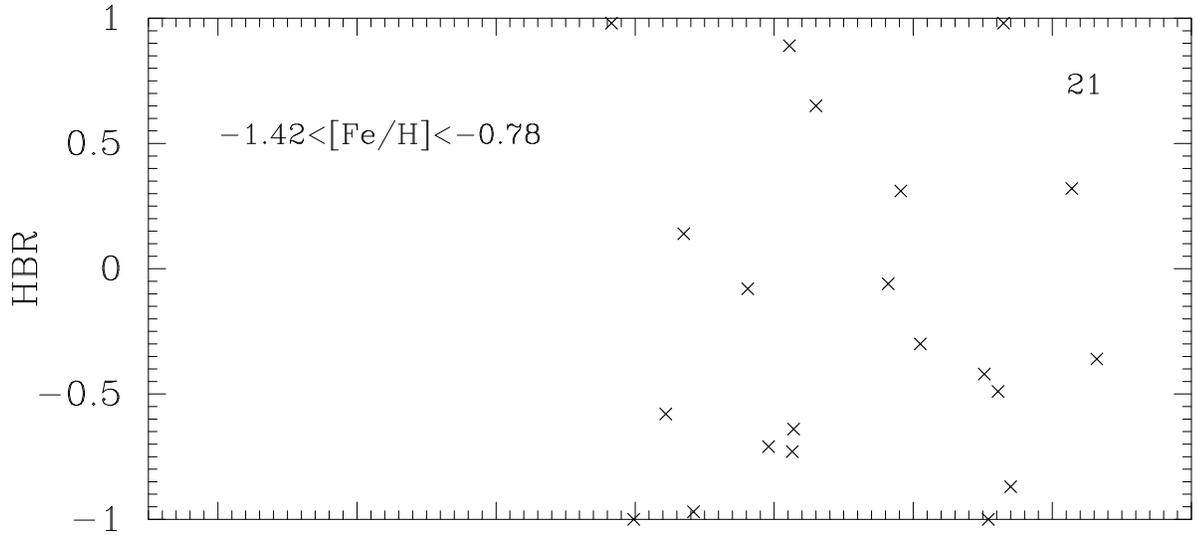
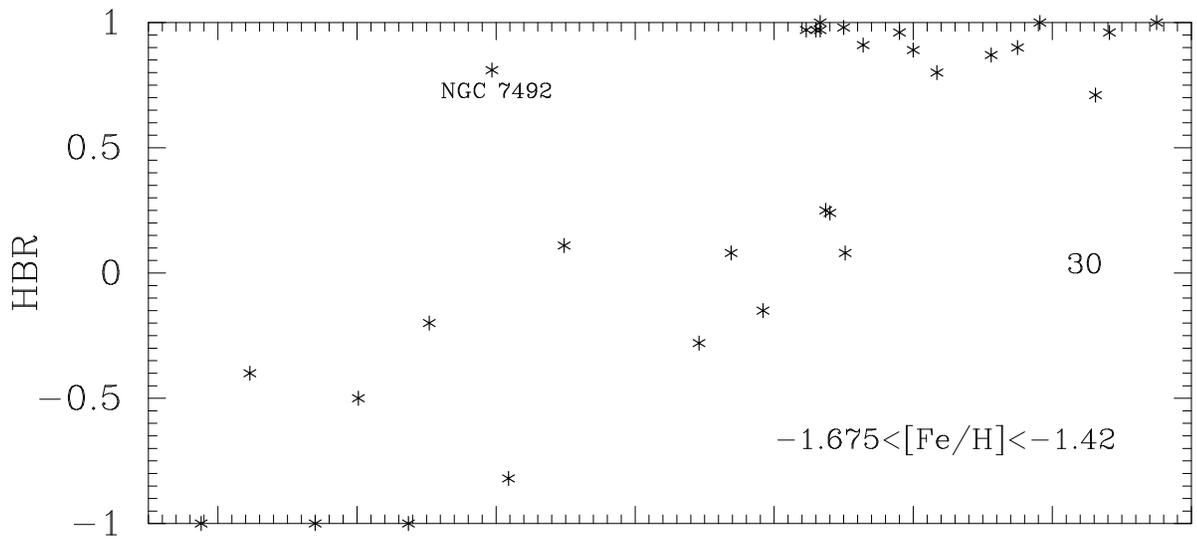
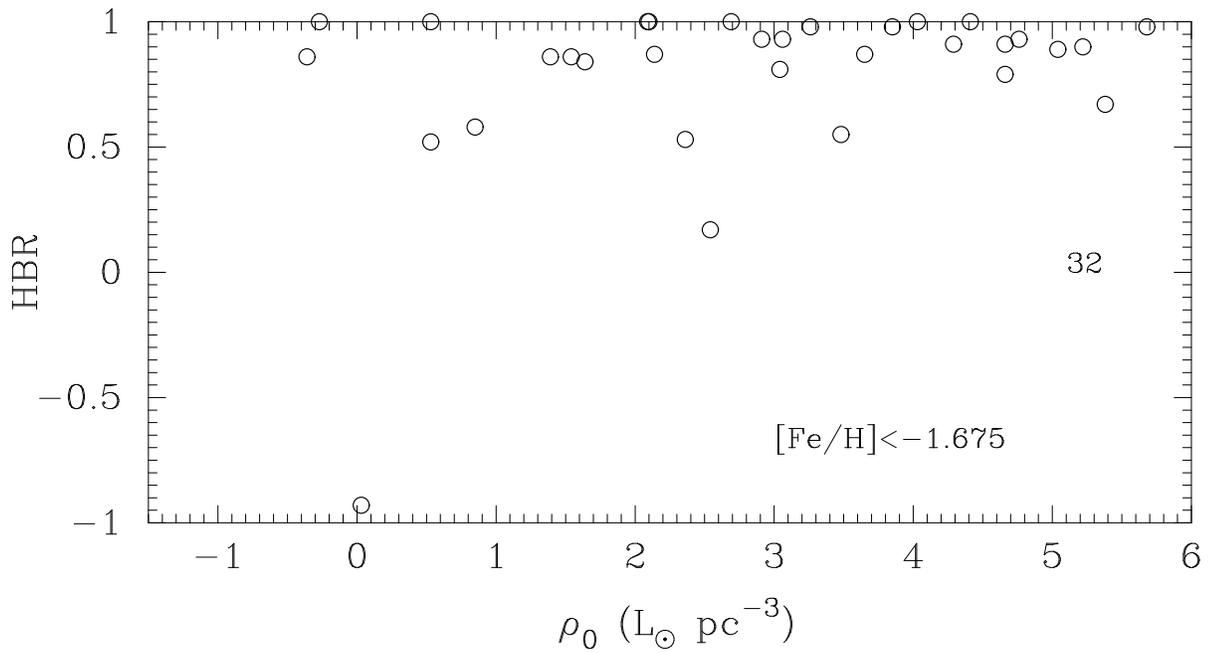
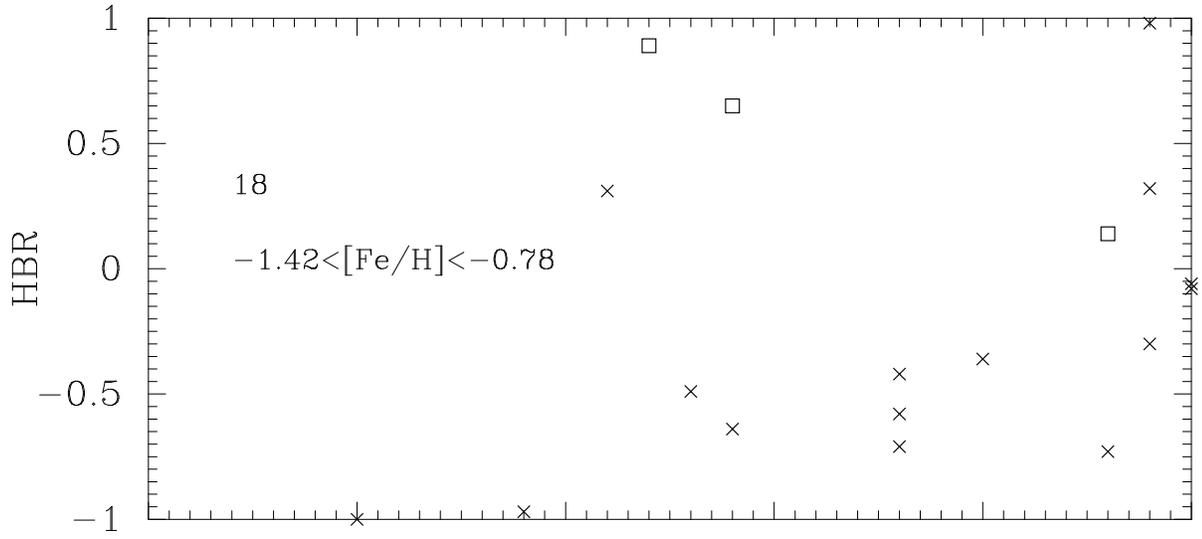
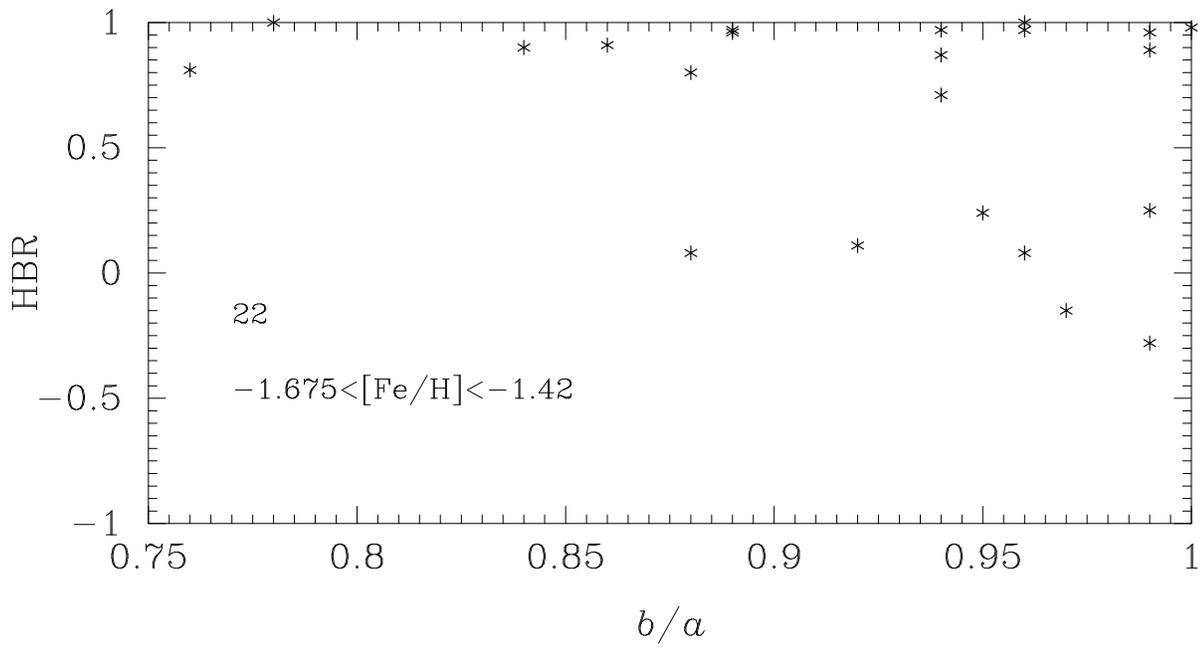
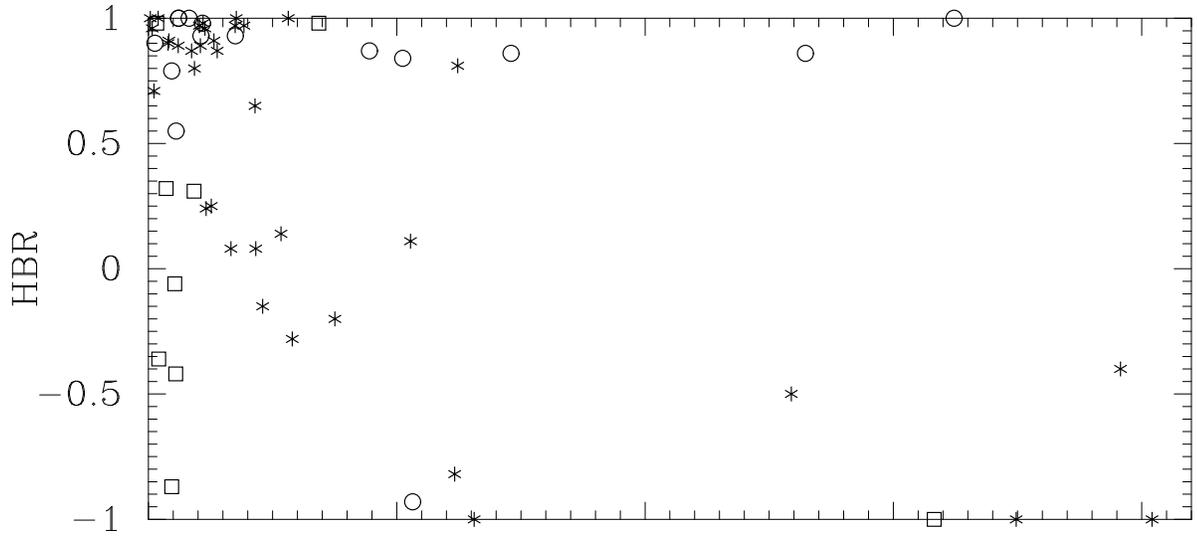
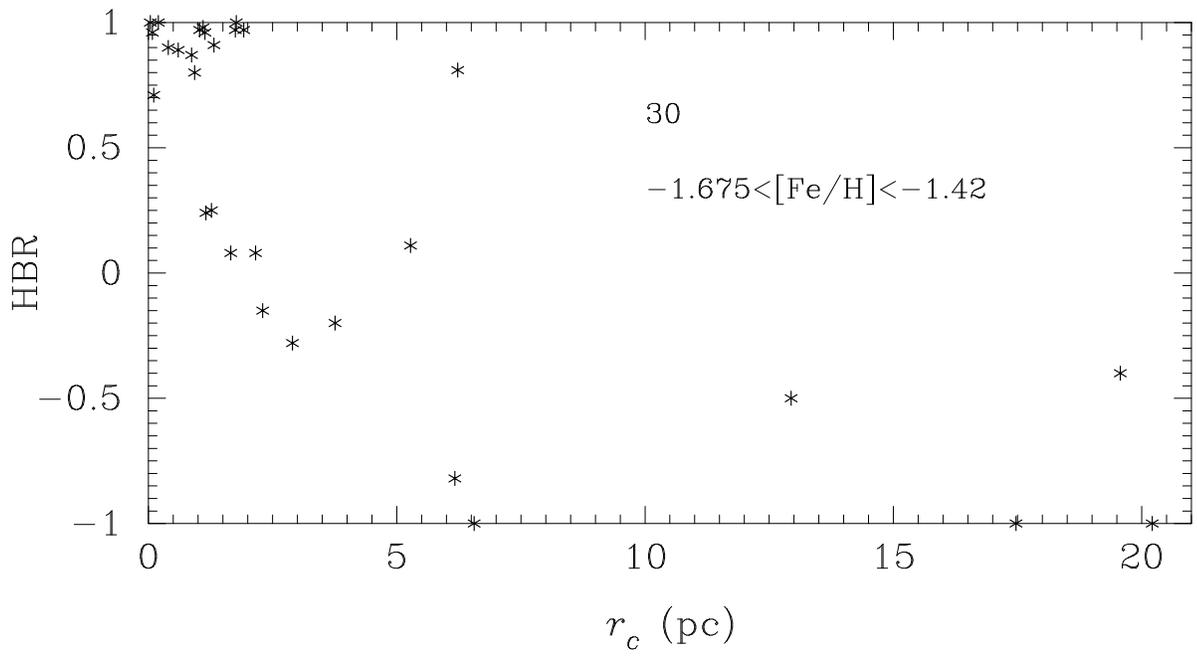
\end{document}